\documentclass[aps,prx,comment,twocolumn,superscriptaddress,amsmath,amssymb]{revtex4-2}
\usepackage{graphicx}
\usepackage{dcolumn}
\usepackage{bm}
\usepackage{color}
\usepackage{hyperref}
\usepackage[T1]{fontenc}
\usepackage{braket}

\begin{document}

\title{Comment on ``Spin-1/2 Kagome Heisenberg Antiferromagnet: Machine Learning Discovery of the Spinon Pair-Density-Wave Ground State''}

\author{Helia Kamal}
\affiliation{Department of Physics, Harvard University, Cambridge, MA 02138, USA}
\affiliation{Harvard Quantum Initiative, Cambridge, MA 02138, USA}
\author{Dominik Kufel}
\affiliation{Department of Physics, Harvard University, Cambridge, MA 02138, USA}
\affiliation{Harvard Quantum Initiative, Cambridge, MA 02138, USA}
\author{DinhDuy Vu}
\affiliation{Department of Physics, Harvard University, Cambridge, MA 02138, USA}
\affiliation{Harvard Quantum Initiative, Cambridge, MA 02138, USA}
% \author{Lode Pollet}
% \affiliation{Department of Physics and Arnold Sommerfeld Center for Theoretical Physics (ASC), Ludwig-Maximilians-Universität München, Theresienstrasse 37, München D-80333, Germany}
% \affiliation{Munich Center for Quantum Science and Technology (MCQST), Schellingstrasse 4, D-80799 München, Germany}
\author{Chris R. Laumann}
\affiliation{Department of Physics, Harvard University, Cambridge, MA 02138, USA}
\affiliation{Department of Physics, Boston University, Boston, MA 02215, USA}
\author{Norman Y. Yao}
\affiliation{Department of Physics, Harvard University, Cambridge, MA 02138, USA}
\affiliation{Harvard Quantum Initiative, Cambridge, MA 02138, USA}

\date{\today}

\begin{abstract}

A recent article [Phys. Rev. X 15, 011047 (2025)] utilizes group equivariant convolutional neural networks to study the ground state of the kagome Heisenberg antiferromagnet.
On the largest finite-size cluster studied to date ($N=108$), the authors report variational energies significantly lower than other numerical methods, including state-of-the-art density matrix renormalization group (DMRG) calculations. 
In contrast to previous results suggesting a possible spin-liquid ground state, the authors observe a spinon pair-density-wave ground state.
We find that: (i) the reported low energies are artifacts of broken ergodicity in the Metropolis–Hastings sampling since the single-spin-flip update rule utilized by the authors effectively freezes the Markov chains, and (ii) when ergodic sampling is enforced via spin-exchange updates, the neural network converges to energies significantly higher than existing DMRG results, calling the paper's claims into question.
\end{abstract}

\maketitle

Understanding the ground state of the spin-1/2 Kagome Heisenberg antiferromagnet (KHAF), with $SU(2)$-symmetric Hamiltonian
$H=J\sum_{\langle ij\rangle}\vec{S}_{i}\cdot\vec{S}_{j}$,
remains one of the most challenging problems in quantum magnetism. 
Correspondingly, a wide range of numerical techniques have been brought to bear on this problem, including:  exact diagonalization (ED)~\cite{Lauchli2019}, variational Monte Carlo (VMC)~\cite{Ran2007,Iqbal2011,Iqbal2014}, density matrix renormalization group (DMRG)~\cite{Jiang2008,Jiang2012,Yan2011,Depenbrock2012,He2017}, and other tensor network approaches ~\cite{Xie2014,Mei2017,Jiang2019}.
Current consensus points to a spin liquid ground state, although the precise nature of the spin liquid  
remains under active debate. 
%
%Resolving this is difficult even with state-of-the-art DMRG algorithms, as tensor network methods are inherently biased toward low-entanglement states and can struggle to faithfully represent gapless phases or long-range entanglement in two dimensions.

Within the VMC framework, \DJ{}uri\'c et al.~\cite{Duric2025}  employ a group convolutional neural network (G-CNN) ansatz and  claim to discover a ``spinon pair-density-wave'' (PDW) ground state with a variational energy 1.78\% lower than the best established DMRG benchmarks~\cite{Depenbrock2012} on a 108-site cluster (see Fig.~5 of Ref.~\cite{Duric2025}). 
This claim is surprising given that the neural network architecture is relatively shallow~\cite{Westerhout2019, Roth2023} and on smaller clusters ($N=48$) where exact diagonalization benchmarks are available, the G-CNN ansatz does not outperform DMRG: instead, it achieves a relative energy error of $\approx 0.3\%$ with respect to exact diagonalization (Fig.~4 of Ref.~\cite{Duric2025}), whereas DMRG reaches $\approx 0.09\%$.

We also note that 
\DJ{}uri\'c et al.~\cite{Duric2025} use two different Monte Carlo sampling schemes in order to optimize the neural network.
% undertake stochastic gradient descent
At smaller system sizes, they use a spin-exchange update rule and observe energies higher than state-of-the-art DMRG. At larger system sizes (including the reported $N=108$ cluster), they use a single-spin-flip update rule and observe energies ``significantly lower'' than state-of-the-art DMRG.

\begin{figure}[ht]
    \centering
    \includegraphics[width=0.90\columnwidth]{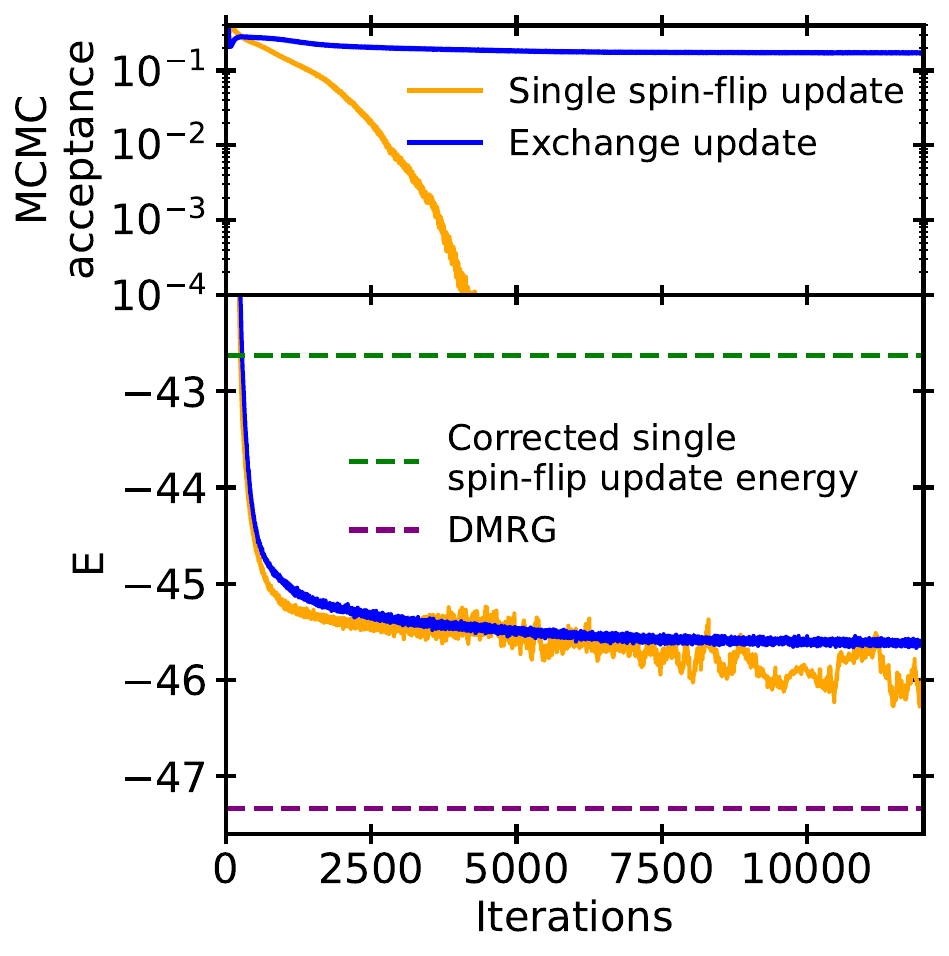}
    \caption{\label{fig:combined} \textbf{(a) Vanishing acceptance rate.} The Markov Chain Monte Carlo (MCMC) acceptance rate during optimization of the $N=108$ cluster (for the $(\Gamma, \chi_0)$ symmetry sector)
     for the single-spin-flip update rule (orange) rapidly collapses to zero (it is exactly zero beyond 5000 iterations), indicating frozen Markov chains. In contrast, the exchange update rule (blue) maintains a finite acceptance rate of approximately $0.2$. Similar behavior was observed on an $N = 48$ cluster. \textbf{(b) Variational energy convergence on the $N=108$ cluster.} We compare two MCMC update rules while keeping the NQS architecture and optimization fixed: (i) optimization using the local spin-flip update rule (orange), as in Ref.~\cite{Duric2025}, yields noisy convergence to a lowest energy of approximately $E \approx -46.2$; (ii) optimization using the exchange update rule (blue) converges stably but to a higher energy, $E \approx-45.6$. As a further test, we take the ansatz parameters optimized using the local spin-flip update rule and re-evaluate the energy using the exchange update rule. The re-evaluated energy (green dashed line) increases to $E \approx -42.6$, confirming that the lower energies obtained via the local spin-flip update are a sampling artifact. 
     The energy computed via DMRG  $E \approx -47.3$ is shown as the black solid line. }

\end{figure}

Given the significance of a new ground-state candidate for the KHAF, it is important to verify the numerical stability of these results. 
Our main findings are as follows:

\begin{enumerate}

    \item The single-spin-flip update rule used for the 108-site cluster is incompatible with an important symmetry of the system, namely, the conservation of total magnetization, $S^z_{\mathrm{tot}}$.
    This results in non-ergodic sampling. 
    As the neural network learns to concentrate around the physical $S^z_{\mathrm{tot}}=0$ sector, the acceptance probability for single-spin-flip updates collapses, leading to frozen Markov chains [Fig.~\ref{fig:combined}(a)].
    
    \item For the largest $N=108$ cluster, when the neural quantum state ansatz is properly optimized using a sampler based on spin-exchange updates, the variational energy converges stably but remains $\approx 3.5\%$ higher than the DMRG benchmark [Fig.~\ref{fig:combined}(b)].
    We note that this behavior is 
    consistent with the method's performance on smaller clusters (also using the spin-exchange update).
\end{enumerate}

\noindent We thus conclude that the reported ``record-low'' energies in Ref.~\cite{Duric2025} are artifacts of broken ergodicity, casting doubt on the  physical interpretation.

\vspace{2mm}

\textbf{Variational Monte Carlo and sampling}--- 
Within VMC, sampling plays a central role in evaluating the loss function (often taken to be the variational energy), gradients, and expectation value of observables. Specifically, the energy expectation value, $E(\vec{\alpha}) = \langle \psi_{\vec{\alpha}} | H | \psi_{\vec{\alpha}} \rangle / \langle \psi_{\vec{\alpha}} | \psi_{\vec{\alpha}} \rangle$,
of a wave-function ansatz parameterized by $\vec{\alpha}$ is computed as the statistical average of the local energy, $E_{\textrm{loc}} (\vec{\alpha}, \vec{\sigma}) = \frac{\sum_{\vec{\sigma}^\prime}  \psi_{\vec{\alpha}}({\vec{\sigma}^\prime}) H_{\vec{\sigma}\vec{\sigma}^\prime}}{\psi_{\vec{\alpha}}(\vec{\sigma})}$, with respect to the Born probability distribution $P_{\vec{\alpha}}(\vec{\sigma}) \propto |\psi_{\vec{\alpha}}(\vec{\sigma})|^2$:
\begin{equation}
        E(\vec{\alpha}) = \sum_{\vec{\sigma}} P_{\vec{\alpha}}(\vec{\sigma}) E_{\textrm{loc}}(\vec{\alpha}, \vec{\sigma}) \approx \frac{1}{N_{s}} \sum_{i = 1}^{N_{s}} E_{\textrm{loc}}(\vec{\alpha}, \vec{\sigma_i}).
\end{equation}
The set of samples $\{ \vec{\sigma}_i \}$ is generated by a Markov Chain Monte Carlo (MCMC) algorithm such as Metropolis-Hastings, where new spin configurations are proposed according to a transition rule---typically a single spin flip (local update) or exchanging two oppositely aligned spins (exchange update)---and accepted with probability
\begin{equation}
    P_{\textrm{acc}}(\vec \sigma \rightarrow \vec \sigma ^\prime) = \min \left(1, \frac{|\psi_{\vec{\alpha}}(\vec{\sigma}^\prime)|^2}{|\psi_{\vec{\alpha}}(\vec{\sigma})|^2} \right).
\end{equation}
In practice, the choice of transition rule strongly influences the efficiency and statistical accuracy of the sampling, and may lead to non-ergodic behavior if incompatible with wavefunction symmetries.

%%%As aforementioned, the KHAF Hamiltonian exhibits a $U(1)$ symmetry corresponding to the conservation of total magnetization, $\hat{S}^z_{\mathrm{tot}}$. 

The KHAF Hamiltonian is fully $SU(2)$ symmetric and its ground state lies in a sector with fixed total magnetization, $\hat{S}^z_{\mathrm{tot}}$. 
Consequently it is natural to initialize the Markov chains in the target symmetry sector and use a magnetization-preserving transition rule, namely the exchange update;
%
%\textcolor{red}{As aforementioned, the KHAF Hamiltonian is fully $SU(2)$ symmetric; in NQS, however, it is straightforward to impose only the $U(1)$ subgroup corresponding to the conservation of total magnetization, $\hat{S}^z_{\mathrm{tot}}$.}
%Consequently, an ansatz trained to minimize the energy naturally suppresses amplitudes for configurations outside the target symmetry sector ($S^z_{\mathrm{tot}}=0$). (this sentence is not relevant here. We have a similar sentence later on where it becomes relevant)
%\textcolor{red}{Consequently it is natural to initialize the Markov chains in the target symmetry sector and use a magnetization-preserving transition rule, namely the exchange update;} 
this is done in Ref.~\cite{Duric2025} for the smaller $N=48$ cluster. 
However, for the larger $N=108$ cluster, Ref.~\cite{Duric2025} instead employs the local update rule, stating that ``the best results are found by sampling within the whole Hilbert space with samples generated by flipping a spin at a random lattice site.''

We find that this choice leads to a vanishing acceptance ratio, frozen Markov chains, and consequently, spurious energy estimates (Ref.~\cite{Duric2025} does not report any acceptance ratios for MCMC). 
This can be understood as follows: as the neural network learns to project the wavefunction onto the $S^z_{\mathrm{tot}} = 0$ symmetry sector, the probability for any configuration with $S^z_{\mathrm{tot}} \neq 0$ is suppressed to near zero. Consequently, any single spin-flip proposal $\vec{\sigma}\to\vec{\sigma}'$ that changes the total magnetization by $\pm 1$ yields a negligible Metropolis-Hastings acceptance ratio,
\begin{equation}
    \frac{|\psi_{\vec{\alpha}}(\vec{\sigma}^\prime)|^2}{|\psi_{\vec{\alpha}}(\vec{\sigma})|^2} \to 0.
\end{equation}
 when $\vec{\sigma}$ lies in the $S^z_{\mathrm{tot}} = 0$ sector. As a result, the Markov chains effectively freeze on a specific configuration and fail to explore the physical Hilbert space.

 \vspace{2mm}

\textbf{Non-ergodicity in the local sampling scheme}--- We attempt to reproduce the results in Ref.~\cite{Duric2025} for the 108-site cluster and utilize the same G-CNN architecture and training scheme (see Appendix). 
We work in the $(\Gamma, \chi_0)$ symmetry sector.

% \textcolor{red}{The symmetry sector we choose to fix the GCNN is $(\Gamma, \chi_0)$ instead of the sector $(M, \chi_0)$ where the authors of Ref.~\cite{Duric2025} claimed ``record-low'' energy (we will justify this choice later in this section). Note that even for the sector $(\Gamma, \chi_0)$, the energy reported in Ref.~\cite{Duric2025} approached the DMRG value.}
%

% \begin{figure}[t]
%     \centering
%     \includegraphics[width=0.8\columnwidth]{Energy.pdf}
%     \caption{\label{fig:energy} \textbf{(b) Variational energy convergence on the $N=108$ cluster.} We compare two MCMC update rules while keeping the NQS architecture and optimization routine fixed: (i) optimization using the local update rule (orange), as in Ref.~\cite{Duric2025}, yields noisy, artificially low energies $\sim -46.2$; (ii) optimization using the exchange update rule (blue) converges stably but to a higher energy $\sim -45.6$. To validate (i), we take the ansatz parameters optimized under local updates and reevaluate the energy using exchange sampling. The corrected estimate (green dashed line) increases to $\sim -42.6$, confirming that the low energies obtained with local updates are sampling artifacts. The energy computed via DMRG  $\sim -47.3$ is shown as the black solid line.}
% \end{figure}

When employing the local spin-flip update rule, we observe a rapidly vanishing MCMC acceptance rate [Fig.~\ref{fig:combined}(a)]. 
This leads to statistical noise in the energy convergence [Fig.~\ref{fig:combined}(b)], consistent with the jagged convergence curves reported in Fig.~5 of Ref.~\cite{Duric2025}.
%This observation explains the jagged energy convergence reported in Fig.~5 of Ref.~\cite{Duric2025}, in contrast to the smoother convergence for the $N=48$ cluster (Fig.~4 of Ref.~\cite{Duric2025}), where the exchange update was used.
%We confirm that this behavior is robust across system sizes, $N = 48$ and $N = 108$, and different space group symmetry sectors.
%
To further characterize the ergodicity of the MCMC dynamics, we quantify the chain mixing via the split-$\hat{R}$ Gelman--Rubin statistic for the energy observable, where $\hat{R}\approx 1$ indicates convergence, while $\hat{R} \approx \sqrt{2}$ indicates completely frozen chains. 
When the ansatz is trained using the local update rule, we find $\hat{R}\approx 1.39$, consistent with freezing. 

As a further test, we take the ansatz trained using the spin-flip update rule and re-evaluate its energy using the exchange rule, where we obtain $\hat{R}\approx 1.03$. 
Doing so, we find an energy which is significantly higher [green dashed line, Fig.~\ref{fig:combined}(b)] than the energy estimated during the original training process.

\vspace{2mm}

\textbf{Ground-state optimization with ergodic sampling}---Repeating the numerical simulations on the 108-site cluster with the exchange update rule yields a stable, finite acceptance rate of $\approx 20\%$, short autocorrelation times, and $\hat{R}\approx 1.03$, confirming that the Markov chains are well-mixed and ergodic [Fig.~\ref{fig:combined}(a)]. 
Under these sampling conditions, the variational energy exhibits smooth convergence to a value higher than that obtained under local spin-flip updates [Fig.~\ref{fig:combined}(b)]. 
Despite extensive hyperparameter exploration and the implementation of an improved optimization scheme (minSR), the lowest energies obtained via the exchange update rule remain higher than the established DMRG benchmarks for the same cluster. 
%

%This strongly suggests that the specific G-CNN architecture employed lacks sufficient expressive power to faithfully capture the true ground state of the 108-site cluster. While increasing the network depth or width could, in principle, enhance expressivity, doing so incurs  computational costs in both memory and runtime, and further exacerbates the difficulty of navigating the highly nonconvex optimization landscape. 
%
% Taken together, these results indicate that, at present, neural quantum states of this form do not yet provide physical insights beyond those already accessible via state-of-the-art DMRG methods for the KHAF.

%The ergodic curve , whereas the non-ergodic curve becomes increasingly noisy as the acceptance rate collapses. 

 \vspace{2mm}

\textbf{Discussion and conclusion}---While our numerics were performed in the $(\Gamma, \chi_0)$ symmetry sector, using the spin-flip update rule should lead to the same frozen Markov chains in any other symmetry sector, including $(M, \chi_0)$ where Ref.~\cite{Duric2025} found the lowest energy state. 
We note that the same authors have recently applied closely related methods to a study of the quantum state associated with the $1/9$ magnetization plateau in the KHAF~\cite{Duric2025plateau}.
Our findings underscore the difficulty of the KHAF problem and highlight the need for rigorous sampling diagnostics in machine-learning studies of quantum many-body systems.

\vspace{2mm}

\textbf{Acknowledgements} We thank Attila Szabó, Lode Pollet and Jack Kemp for discussions and feedback on the Comment.

\section*{Appendix: numerical methods}

Our simulations use a group convolutional neural network (G-CNN) ansatz with 6 layers and 6 feature maps per layer on a $4\times 4\times 3=48$-site cluster, and 4 layers and 4 feature maps per layer on a $6\times 6\times 3=108$-site cluster, following Ref.~\cite{Duric2025}. The architecture uses the scaled exponential linear unit (SELU) nonlinearity, applied separately to the real and imaginary parts of the feature maps \cite{Roth2021, Roth2023}. To produce the probability amplitude $\psi(\sigma)$ for an input spin configuration, the output layer projects the sum of exponentiated feature maps onto a specific irreducible representation of the space group characterized by characters $\{\chi_{g}\}_{g\in G}$ \cite{Roth2021, Roth2023}. The simulations were performed in NetKet \cite{Netket}. The energy is evaluated using $2^{13}$ samples, divided into $2^9$ chains.

\end{document}